\documentclass{iau}
\usepackage{graphicx}

\title[Simulated fluxes of oscillating tori around Kerr black holes] %% give here short title %%
{Simulations of flux variability of oscillating\\ accretion fluid tori around Kerr black holes}

\author[Pavel Bakala et al]   %% give here short author list %%
{Pavel Bakala$^1$, Kate{\v{r}}ina Goluchov{\'{a}}$^1$, Eva {\v{S}}r{\'{a}}mkov{\'{a}}$^1$, \\Andrea Kotrlov{\'{a}}$^1$,
Gabriel T\"or\"ok$^1$,\\Frederic H. Vincent$^2$ and Marek A. Abramowicz$^{1,2}$}

\affiliation{$^1$Institute of  Physics, Faculty of Philosophy and Science, \\Silesian University in Opava,
    Bezru\v{c}ovo n\'{a}m. 13, Opava, Czech Republic\\ e-mail: {\tt pavel.bakala@fpf.slu.cz} \\[\affilskip]
$^2$Nicolaus Copernicus Astronomical Center, Warszawa, Poland}

\pubyear{2014}
\volume{313}  %% insert here IAU Symposium No.
\pagerange{119--126}
\jname{Extragalactic jets from every angle}
\editors{F. Massaro, C. C. Cheung, E. Lopez, and A. Siemiginowska, eds.}
\begin{document}

\maketitle

\begin{abstract}
High frequency quasi-periodic oscillations (HF QPOs) are observed in the X-ray power-density spectra (PDS) of several microquasars and low mass X-ray binaries. Many proposed QPO models are based on oscillations of accretion toroidal fluid  structures orbiting in the vicinity of a compact object. We  study  oscillating accretion tori orbiting in the vicinity of a Kerr black hole. We demonstrate that significant variation of the observed flux can be caused by the combination of radial and vertical oscillation modes of a slender, polytropic, perfect fluid, non-self-graviting torus with constant specific angular momentum. We investigate two combinations of the oscillating modes corresponding to the direct resonance QPO model and the modified relativistic precession QPO model .
\keywords{X-rays: binaries, stars: neutron, black hole physics, accretion, accretion disks}
\end{abstract}

We assume an axially-symmetric torus in a hydrostatic equilibrium. We also assume a perfect, polytropic, non-self-gravitating fluid with constant specific angular momentum distribution. The torus is situated on the background of Kerr geometry. The solution of the perturbed Papaloizou-Pringle equation (\cite{pap.pri.1984,Bla.etal.2006}) yields the complete set of discrete eigenfunctions and eigenfrequencies that correspond to different oscillation modes. We consider here the simplest radial and vertical epicyclic oscillation modes whose eigenfunctions are pure functions of the radial and vertical coordinate, and the corresponding eigenfrequencies are linear combinations of the  appropriate epicyclic frequency of a test particle and Keplerian frequency $\nu_{K}$. All frequencies are defined on the radial coordinate of torus centre. The centre of the torus is situated in such a way that the ratio of mode frequencies is equal to 3/2 in accordance with the observations of double peaks HF QPOs in PDS of microquasars and LMXBs (\cite{tor.etal.2007}). The first investigated torus setup combines pure epicyclic axisymmetric vertical and radial modes, where mode frequencies are identical with radial $\nu_{r}$ and vertical $\nu_{\theta}$ epicyclic frequencies, respectively (\cite{abr-klu:2001}).
The second setup combines non-axisymmetric vertical mode with frequency $\nu_{u}=2\nu_{K}-\nu_{\theta}$ and non-axisymmetric radial mode with frequency $\nu_{l}=\nu_{K}-\nu_{r}$. In the Schwarzschild case, such frequency relation corresponds exactly to Relativistic Precession QPO model (\cite{ste-vie:1999}).  In the Kerr case, the deviations from such  model are small (\cite{tor.etal.2010}).  
\begin{figure}[h!]
\centering
\includegraphics[width=0.315\hsize, angle=0]{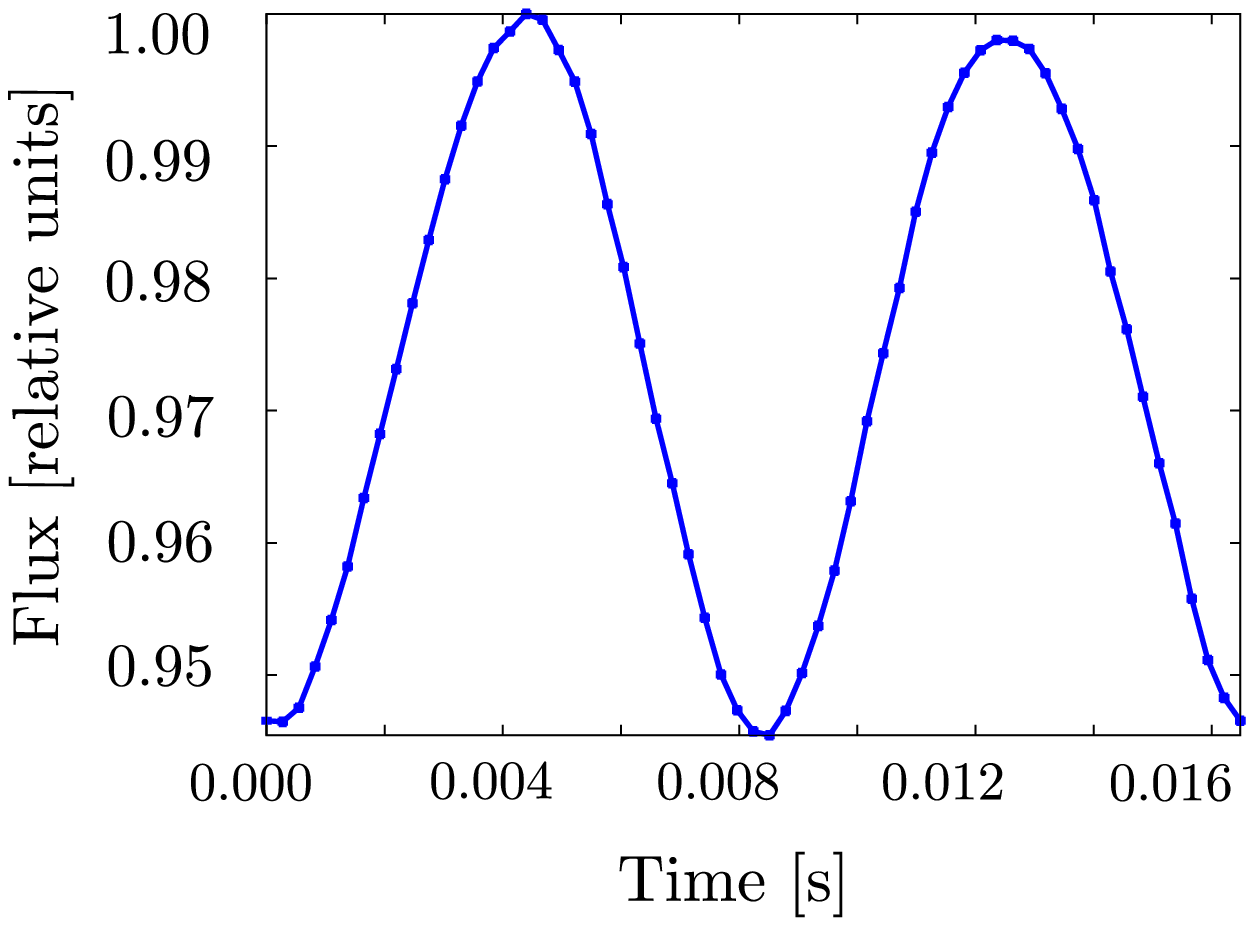}
\includegraphics[width=0.315\hsize, angle=0]{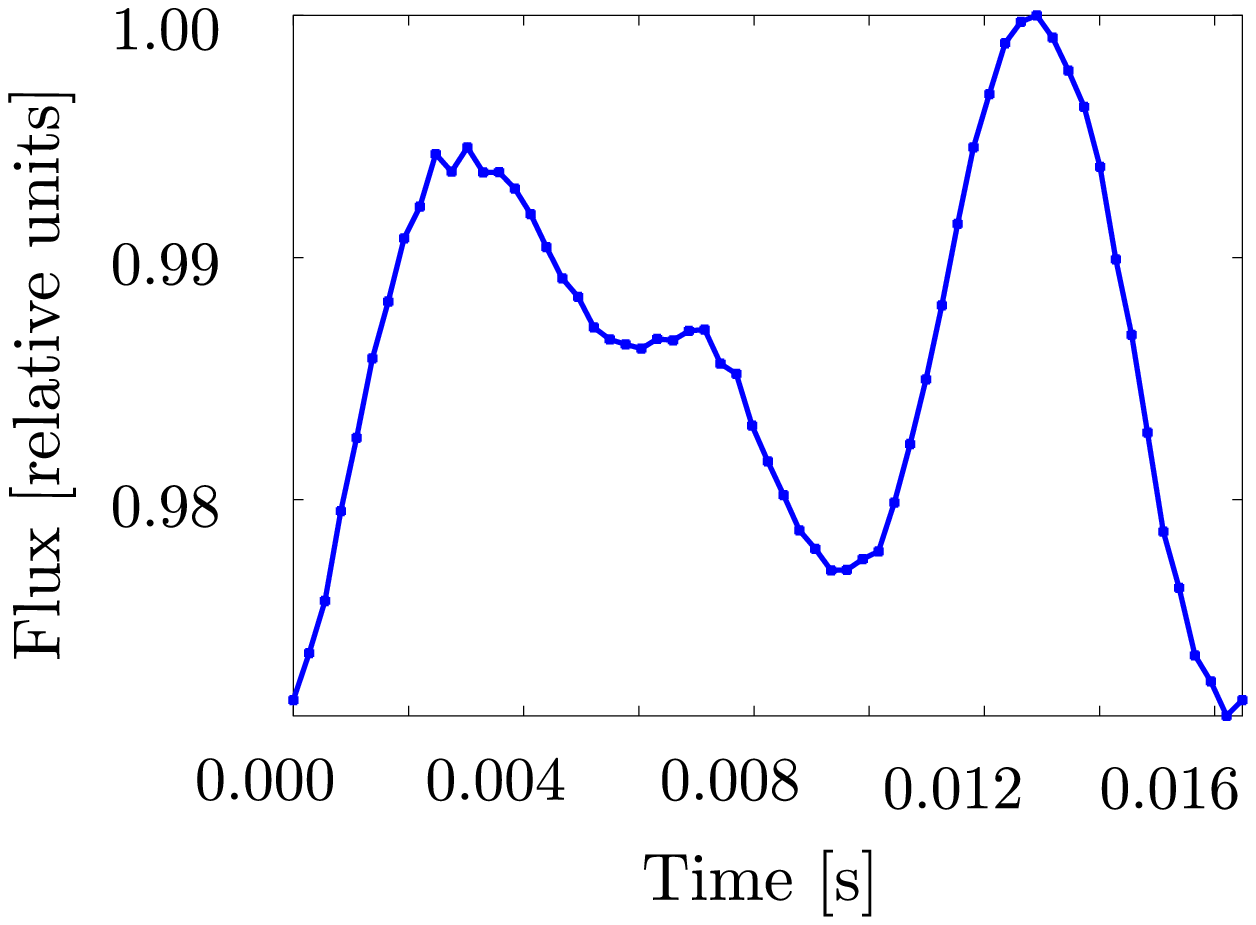}
\includegraphics[width=0.315\hsize, angle=0]{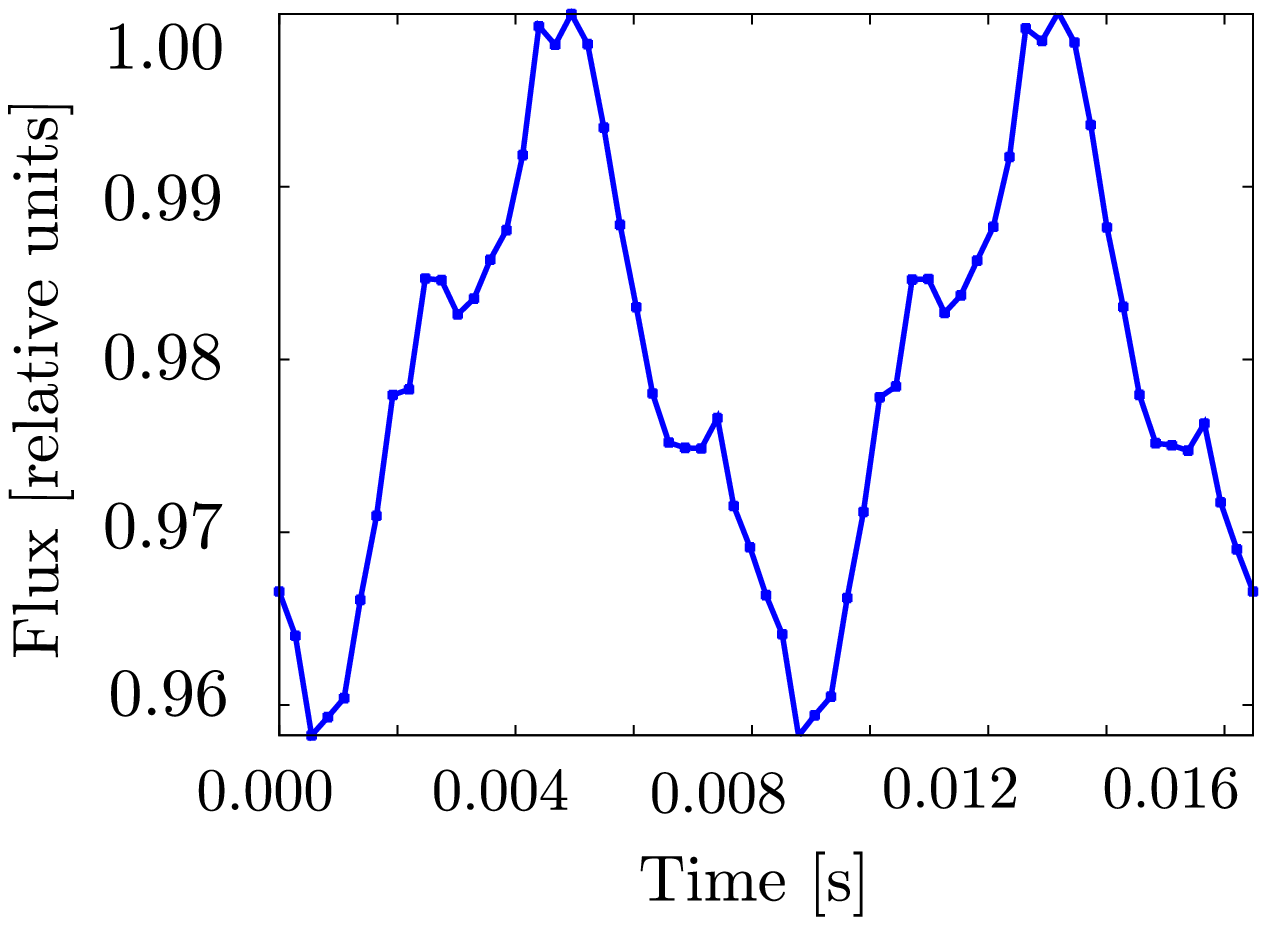}
\includegraphics[width=0.315\hsize, angle=0]{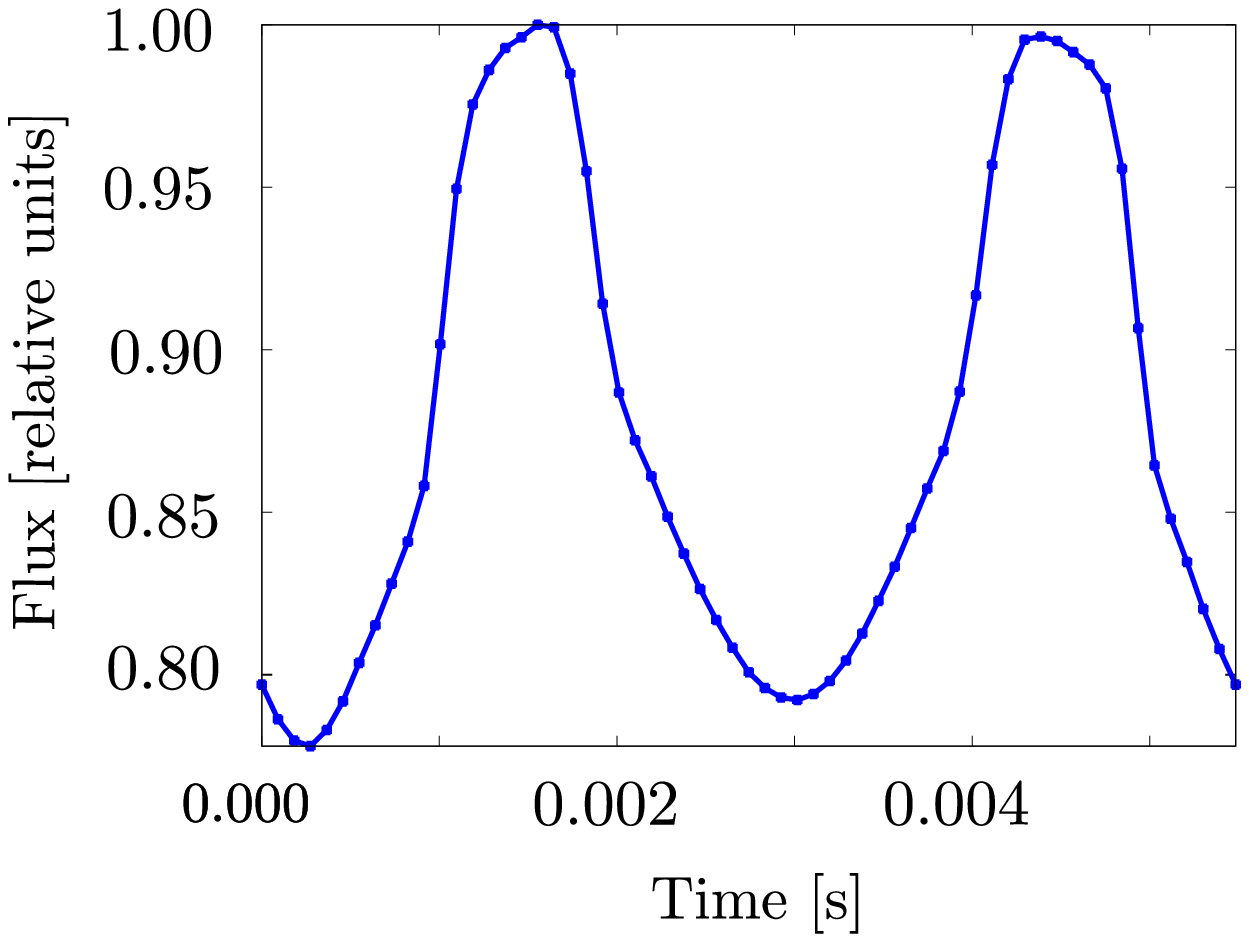}
\includegraphics[width=0.315\hsize, angle=0]{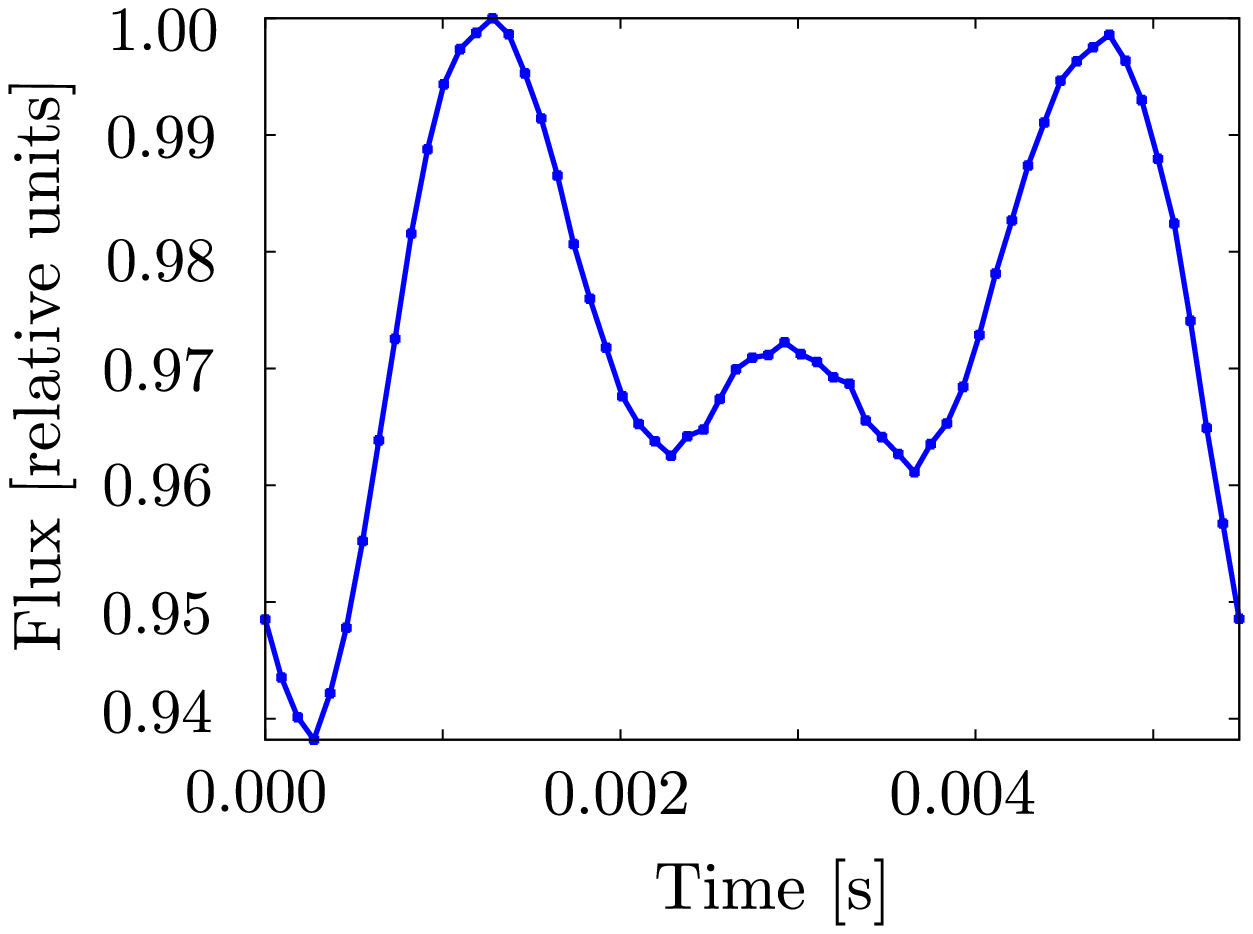}
\includegraphics[width=0.315\hsize, angle=0]{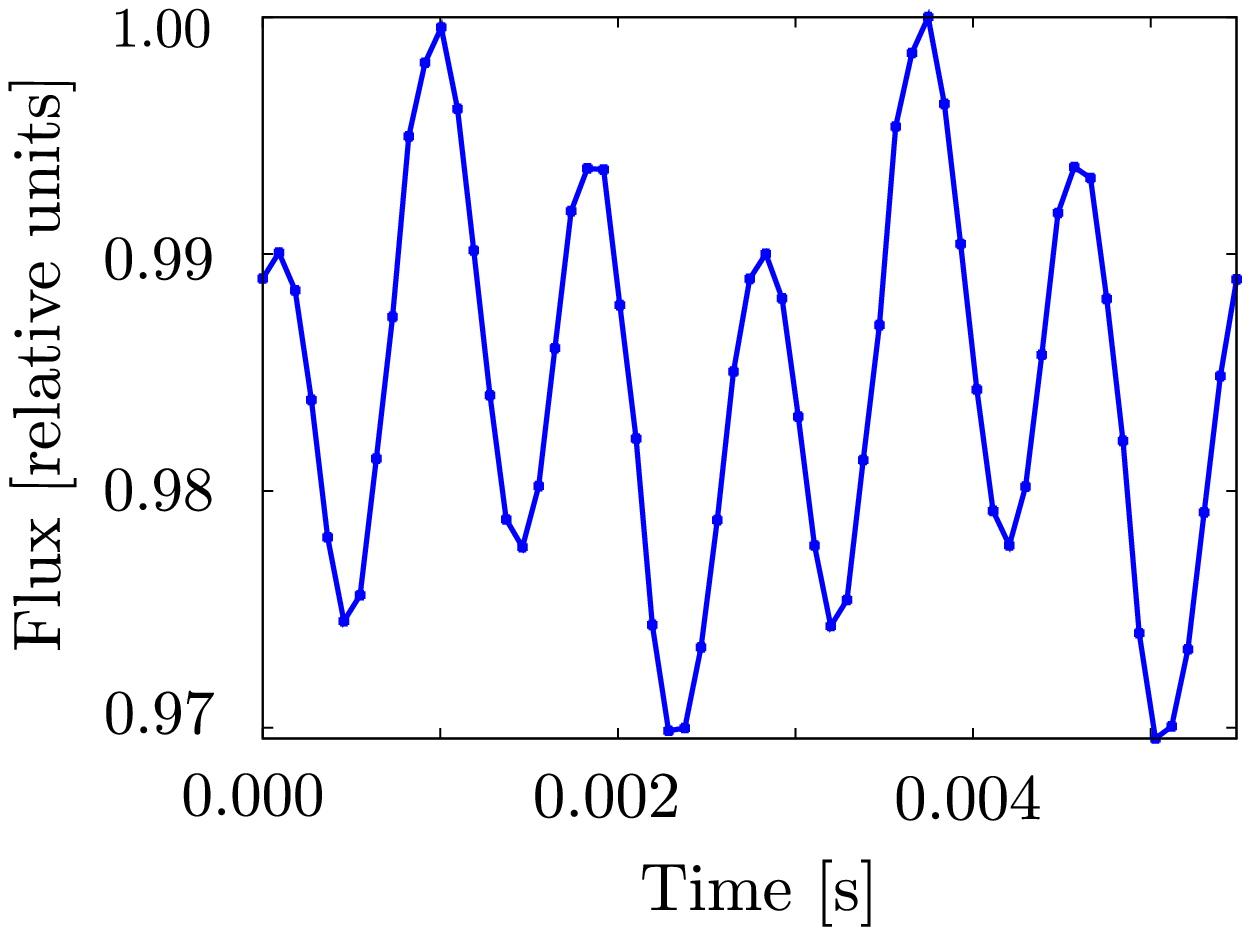}
\noindent
\caption[t]
{Light curves of the oscillating torus with pure epicyclic axisymmetric modes. Plots are constructed for different observer inclination: $\theta_{obs} = 0.01$ (left column), $\theta_{obs} = 1/3 \pi$ (middle column) and $\theta_{obs} = 1/2 \pi$ (right column). The top line corresponds to $a = 0$ and the bottom one to $a = 0.96$.}
\label{figure1}
\end{figure}

\begin{figure}[h!]
\centering
\includegraphics[width=0.315\hsize, angle=0]{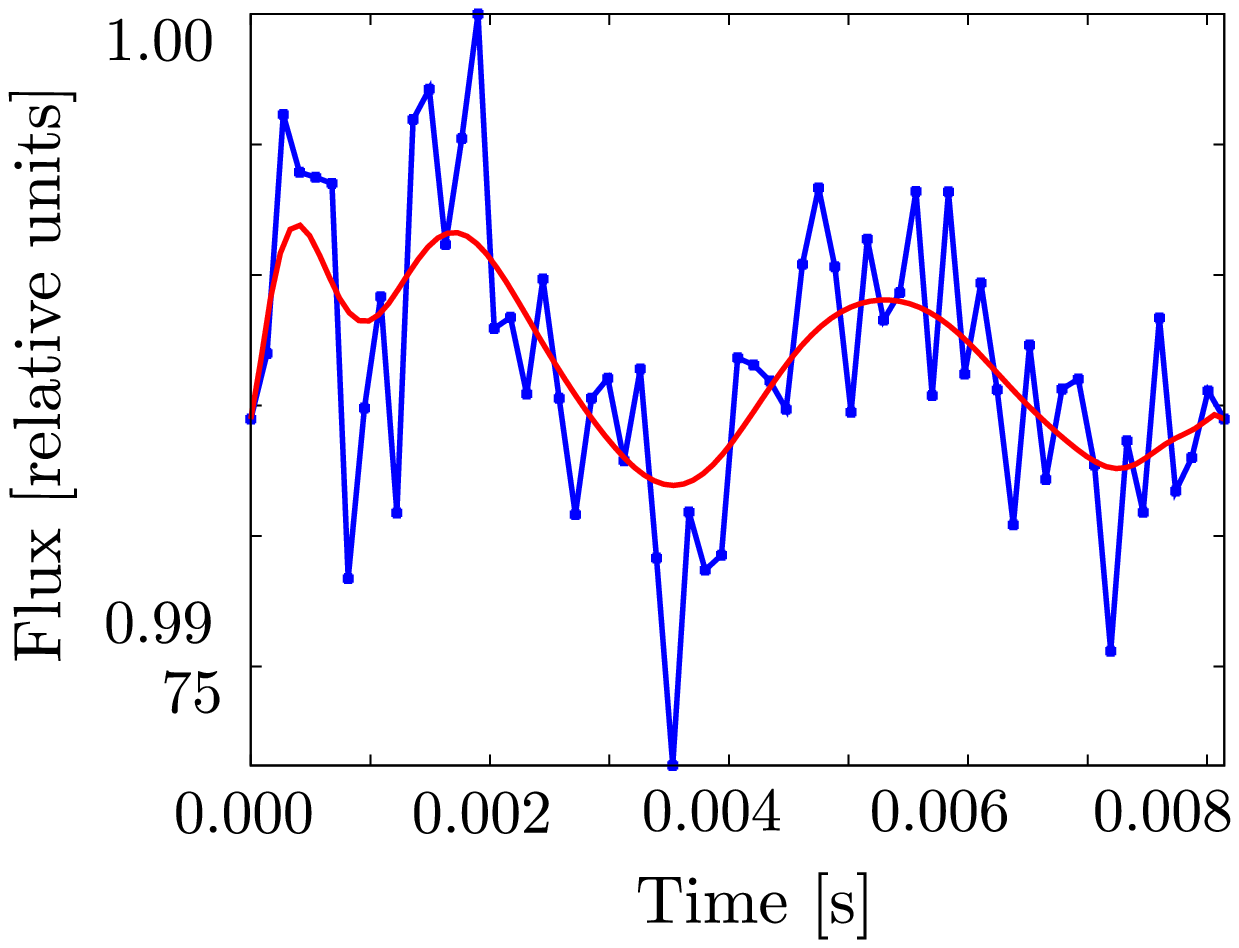}
\includegraphics[width=0.315\hsize, angle=0]{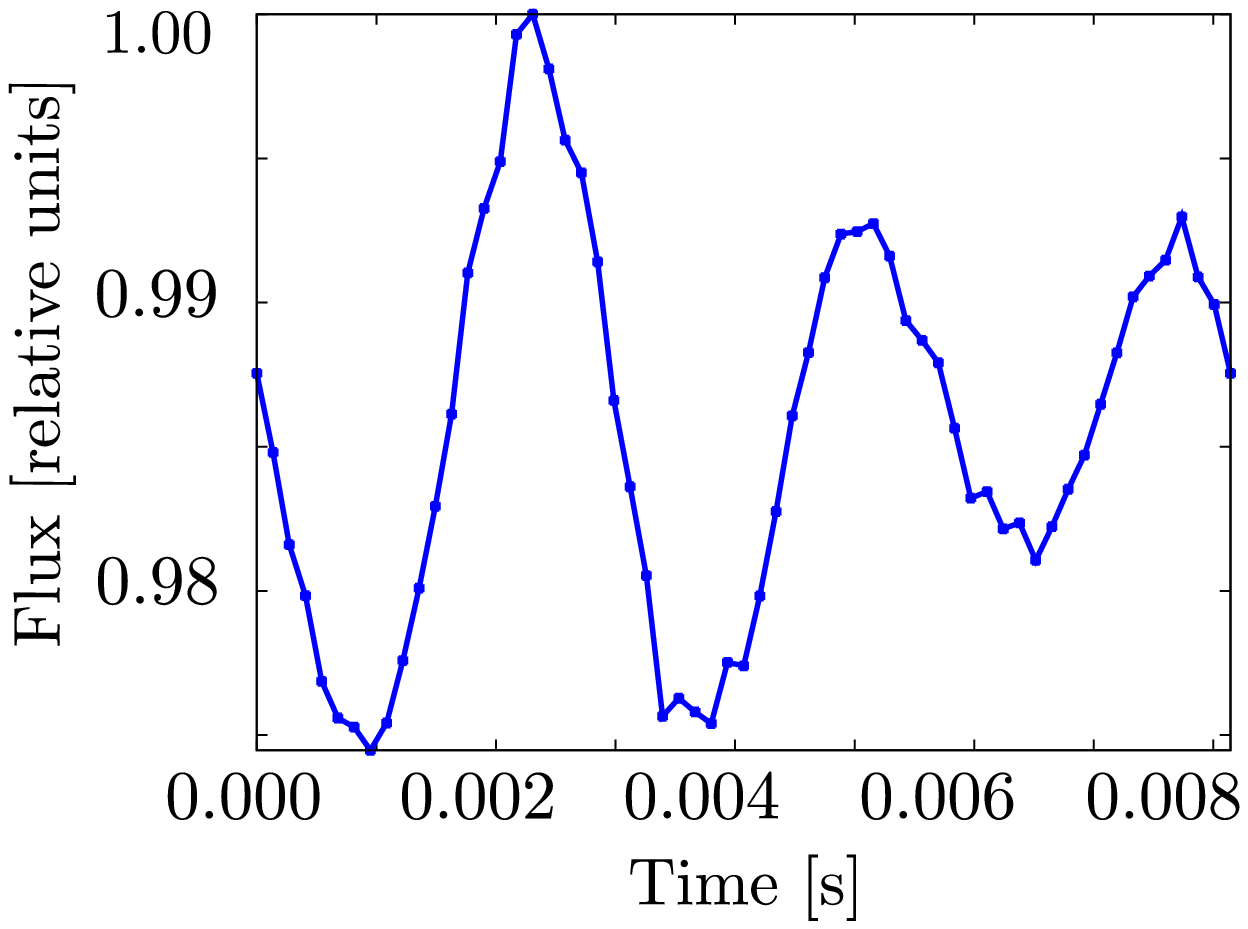}
\includegraphics[width=0.315\hsize, angle=0]{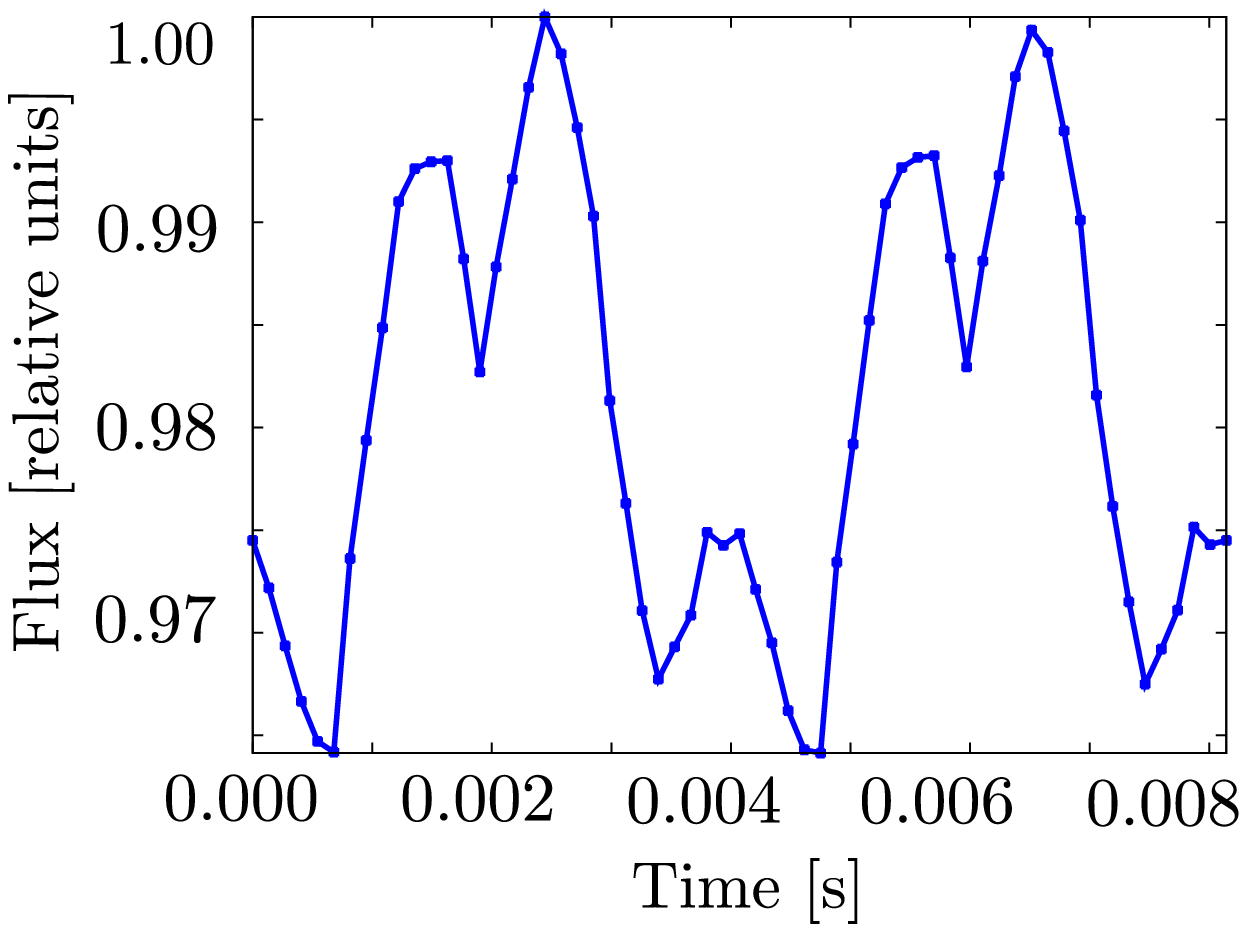}
\includegraphics[width=0.315\hsize, angle=0]{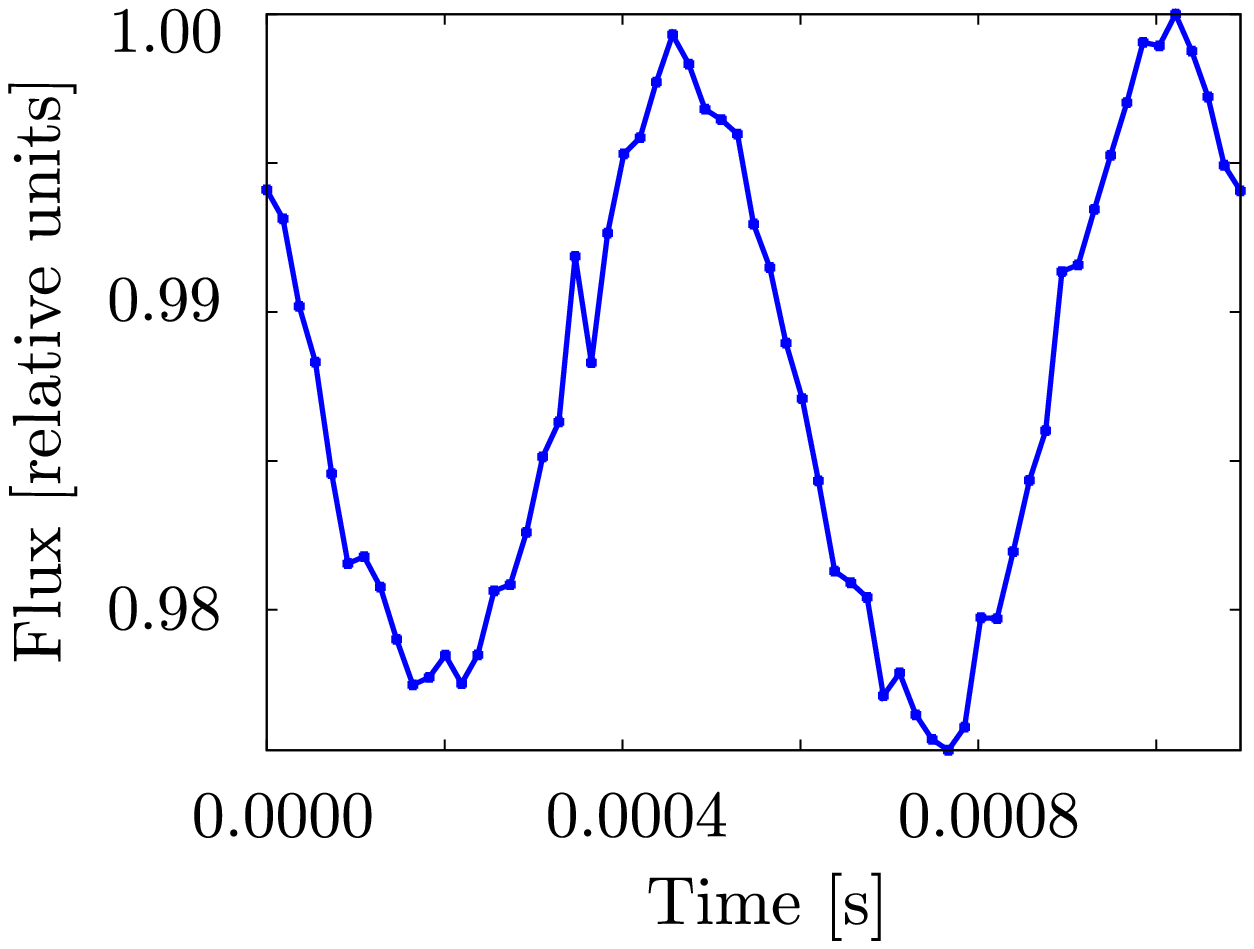}
\includegraphics[width=0.315\hsize, angle=0]{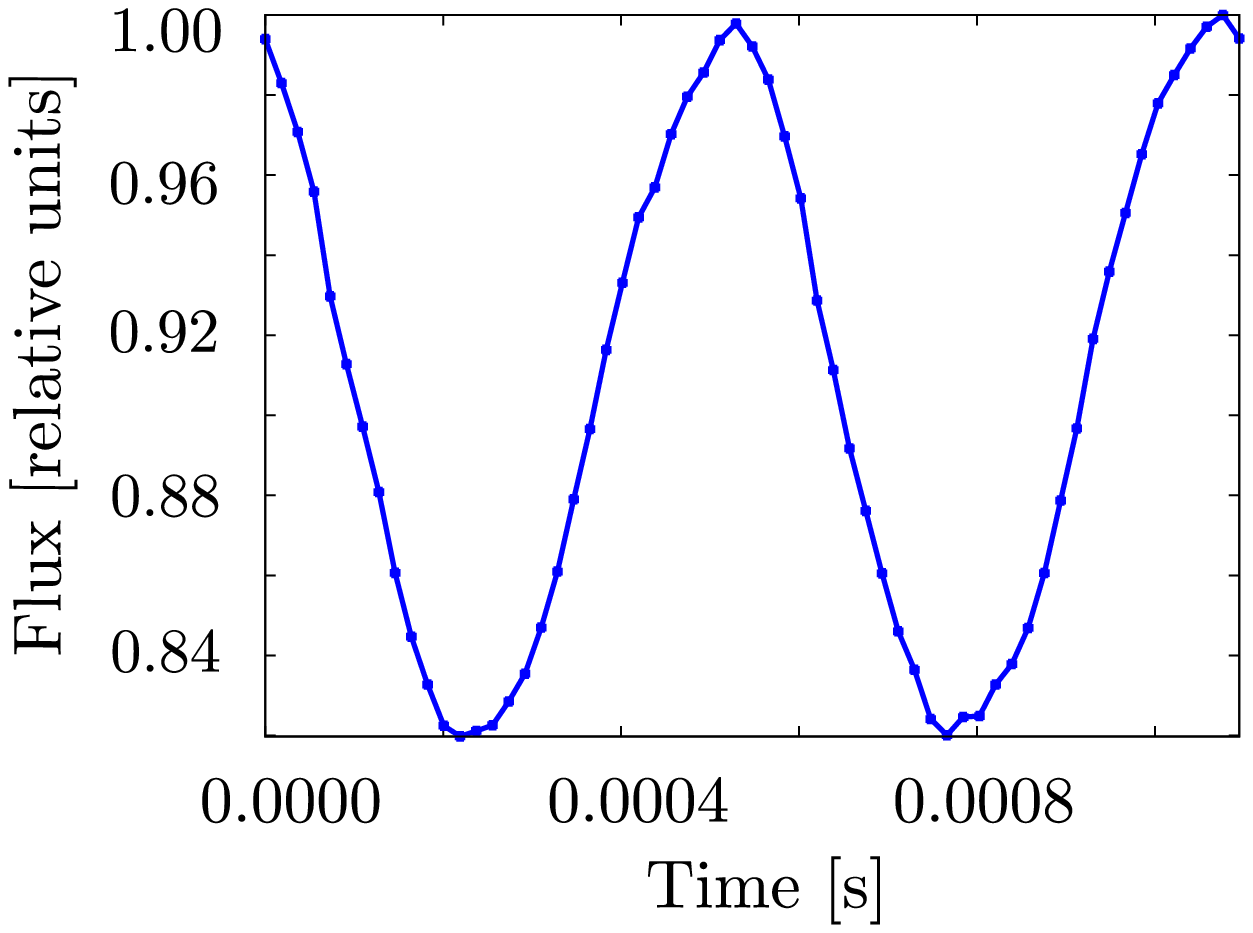}
\includegraphics[width=0.315\hsize, angle=0]{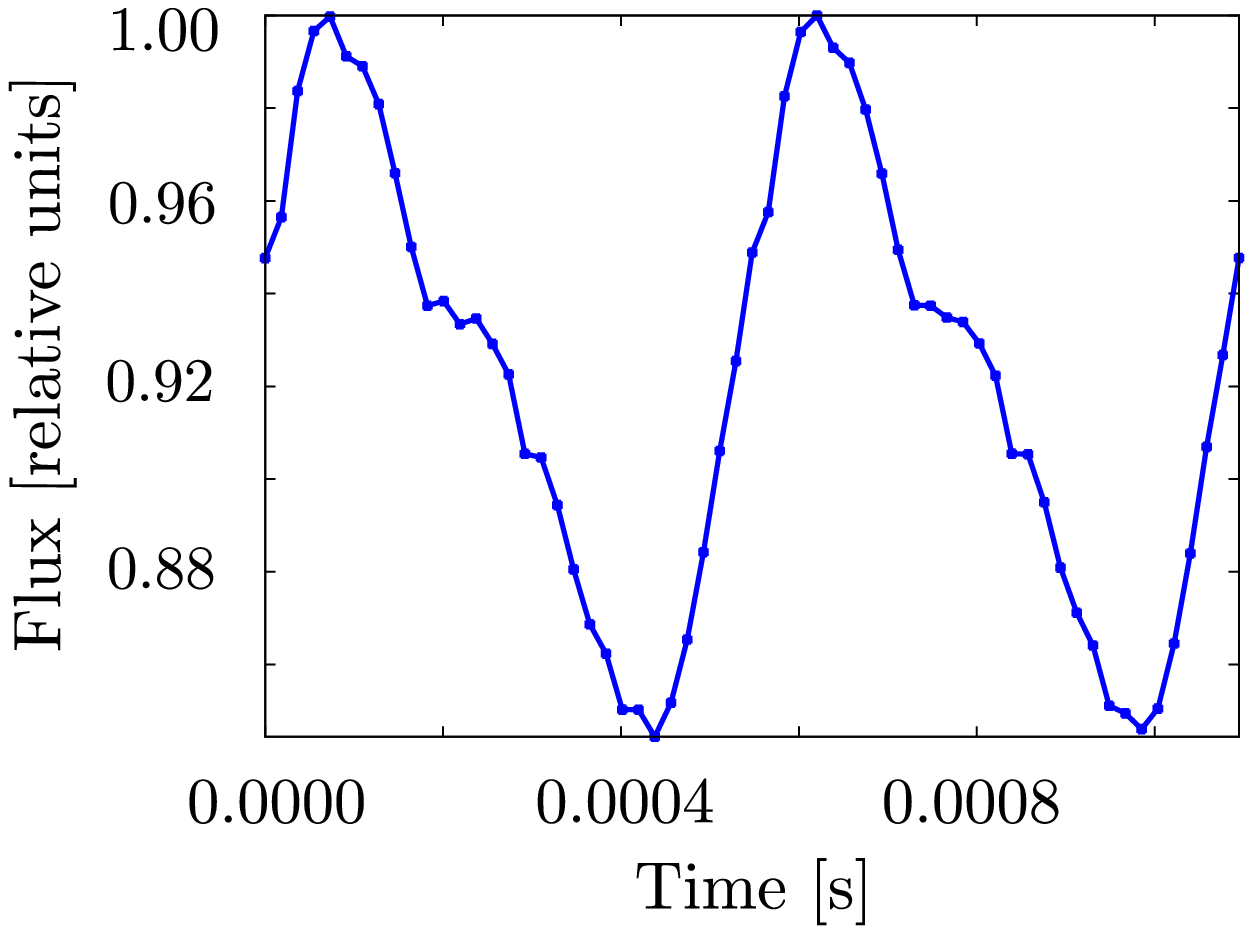}
\noindent
\caption[t]
{Light curves of the Relativistic Precession QPO model like combination of non-axisymmetric modes presented in the same way as in Figure \ref{figure1}.}
\label{figure2}
\end{figure}

\section*{Acknowledgments}
The authors acknowledge the project CZ.1.07/2.2.00/28.0271 "INAP", the Czech grant GA\v{C}R~209/12/P740, and the internal grant of SU Opava, SGS/11/2013.


\begin{thebibliography}{}
%%%%%%%%%%%%%%%%%%%%%%%%%%%%%%%%%%%%%%%%%%%%%%%%%%%%%%%%%%%%%%%%%
\bibitem[Abramowicz \& Klu{\'z}niak(2001)]{abr-klu:2001}{Abramowicz, M. A. \& Klu{\'z}niak, W.} 2001, \textit{A\&A}, v.374, p.L19--L20
%%%%%%%%%%%%%%%%%%%%%%%%%%%%%%%%%%%%%%%%%%%%%%%%%%%%%%%%%%%%%%%%%%%%%%%%%%%%%%%%%%%%%%
\bibitem[Blaes et al. 2006]{Bla.etal.2006}{ Blaes,O. M., Arras, P., Fragile,P. C.} 2006,
\textit{MNRAS},  369,  1235-1252.
%%%%%%%%%%%%%%%%%%%%%%%%%%%%%%%%%%%%%%%%%%%%%%%%%%%%%%%%%%%%%%%%%%%%%%%%%%%%%%%%%%%%%%
\bibitem[Papaloizou \& Pringle (1984)]{pap.pri.1984}{Papaloizou, J. C. B. \& Pringle, J. E.} 1984,
\textit{MNRAS},  208,  721-750.
%%%%%%%%%%%%%%%%%%%%%%%%%%%%%%%%%%%%%%%%%%%%%%%%%%%%%%%%%%%%%%%%%%%%%%%%%%%%%%%%%%%%%%
%%%%%%%%%%%%%%%%%%%%%%%%%%%%%%%%%%%%%%%%%%%%%%%%%%%%%%%%%%%%%%%%%%
\bibitem[Stella \& Vietri (1999)]{ste-vie:1999}{Stella, L. \& Vietri, M.} 1999, \textit{Phys. Rev. Lett}, 82, 17.
%%%%%%%%%%%%%%%%%%%%%%%%%%%%%%%%%%%%%%%%%%%%%%%%%%%%%%%%%%%%%%%%%%%%%%%%%%%%%%%%%%%%%%
\bibitem[T\"{o}r\"{o}k et al. (2007)]{tor.etal.2007}{T\"{o}r\"{o}k, G., Stuchl\'{i}k Z. and Bakala,P.}, 2007,
\textit{Central European Journal of Physics},  5, 457-462.
%%%%%%%%%%%%%%%%%%%%%%%%%%%%%%%%%%%%%%%%%%%%%%%%%%%%%%%%%%%%%%%%%%
\bibitem[T\"{o}r\"{o}k et al. (2010)]{tor.etal.2010}{T\"{o}r\"{o}k, G., Bakala, P., \v{S}r\'{a}mkov\'{a}, E., Stuchl\'{i}k, Z. and Urbanec M.},  2010,
\textit{ApJ}, 714, 748-757.

\end{thebibliography}
\end{document}